\documentclass[prl,nofootinbib,superscriptaddress,twocolumn,showpacs]{revtex4-1}

\usepackage{bm}
\usepackage{amsfonts}
\usepackage[pdftex]{color,graphicx}
\usepackage{amssymb}
\usepackage{amsmath}
\usepackage{amsthm}
\usepackage{hyperref}
\usepackage{times}

\newcommand{\seq}[1]{\langle #1 \rangle}

\newcommand{\JJ}{\mathcal{J}}

\let\oldmarginpar\marginpar
\renewcommand\marginpar[1]{\-\oldmarginpar[\raggedleft\tiny\color{red} #1]%
{\raggedright\tiny #1}}

\graphicspath{{Figures/}}

\begin{document}


\title{Disorder-induced Majorana metal in interacting non-Abelian anyon systems}
	
\author{Chris R. Laumann}
\affiliation{Department of Physics, Harvard University, Cambridge, MA 02138}

\author{Andreas W.W. Ludwig}
\affiliation{Department of Physics, University of California, Santa Barbara, CA 93106}

\author{David A. Huse}
\affiliation{Department of Physics, Princeton University, Princeton, NJ 08544}

\author{Simon Trebst}
\affiliation{Microsoft Research, Station Q, University of California, Santa Barbara, CA 93106}

\date{\today}

\begin{abstract}
We demonstrate that a thermal metal of Majorana fermions forms in a 
two-dimensional system of interacting non-Abelian (Ising) anyons in the presence of moderate disorder.
This bulk metallic phase arises in the $\nu=5/2$ quantum Hall state 
when disorder pins the anyonic quasiparticles. More generally, it naturally occurs for various proposed systems supporting Majorana fermion zero modes when disorder induces the random pinning of a finite density of vortices. This includes all two-dimensional topological superconductors in so-called symmetry class D.  A distinct experimental signature of the thermal metal phase is the presence of bulk heat transport down to zero temperature.
\end{abstract}



\maketitle


Quantum matter with exotic non-Abelian quasiparticle statistics has become a highly sought-after
state of matter \cite{Stern}.
Theoretical proposals suggest the existence of the most elementary incarnation of such a state -- arising
from zero modes of Majorana fermions -- in diverse  two-dimensional (2D) systems such as the fractional quantum Hall liquid at  $\nu=5/2$
filling \cite{FiveHalfState}, p$_x$+ip$_y$ superconductors \cite{ReadGreen}, heterostructures of topological insulators
and superconductors \cite{FuKanePplusIP}, and possibly certain Iridates which effectively realize the Kitaev honeycomb model \cite{Iridates}.
However, despite intense efforts, experimental observation of these states has proven elusive to date.
In this manuscript we point out a distinct collective phenomenon of these states in the presence of disorder --
the formation of a thermal metal manifesting itself in bulk heat transport which may be amenable to a more
direct observation than previously considered experimental signatures such as sophisticated
interferometric setups \cite{Interferometer}.
As we show, this thermal metal phase arises when a finite density of quasiparticles is spatially pinned by disorder,
and  the resulting macroscopic degeneracy -- characteristic of non-Abelian statistics -- hybridizes in the presence of
interactions between the quasiparticles.
For concreteness, we will discuss our results in the context of the $\nu=5/2$ quantum Hall state, but point out that
our results hold for other 2D systems with randomly pinned zero modes of Majorana fermions.

\paragraph{The disordered $\nu=5/2$ quantum Hall state.--}
Since the seminal work by Moore and Read \cite{MooreRead} the quantum Hall state on the $\nu=5/2$
plateau has been a prominent candidate state supporting quasiparticles which obey non-Abelian statistics.
Detuning the magnetic field from the middle of this plateau will introduce a finite density of  charged
quasiparticles (or quasiholes) which for a clean system will naturally form  a triangular Wigner crystal,
as recently observed  for the $\nu=1/3$ quantum Hall state \cite{WignerCrystal}.
Even though pinned, these quasiparticles still carry a fluctuating quantum mechanical degree of freedom,
similar to the spin degree of freedom of an electron in a Mott insulator. It is this degree of freedom, which
encodes the non-Abelianness of the quasiparticle, that is referred to as an Ising anyon $\sigma$.
Similar to a pair of ordinary spin-1/2 moments, which can combine as $1/2 \times 1/2 = 0 + 1$ into a singlet
or triplet state, two Ising anyons can form two distinct collective states
\begin{equation}
	\sigma\times\sigma=1+\psi \,,
	\label{Eq:FusionRule}
\end{equation}
where the state $1$ refers to the vacuum (or trivial) state and $\psi$ is another quasiparticle
excitation.
An immediate consequence of this so-called fusion rule \eqref{Eq:FusionRule} is the formation of a
macroscopic degeneracy of states, which in the case of $N$ ordinary spin-1/2 moments grows like $2^N$
and for the case of Ising anyons grows asymptotically as $\sqrt{2}^N$ for a system of $N$ Ising anyons. 
This macroscopic degeneracy is lifted by interactions between the Ising anyons.
The most elementary example of such a splitting occurs for a
pair of Ising anyons brought into spatial proximity which will naturally split the energies of the two
states on the right-hand side (rhs) in Eq.~\eqref{Eq:FusionRule}, in analogy to the splitting of singlet and triplet
states for a pair of ordinary spin-1/2 moments.
For more than two anyons the many-body problem arising from these
interactions can be formulated \cite{PhysRevLett.98.1604096}
in terms of pairwise projectors $\Pi_{jk}$ \cite{FootnotePairwiseInteractions}
energetically favoring one of the two states on the rhs of Eq.~\eqref{Eq:FusionRule}
\begin{equation}
	\label{eq:ham_lattice}
	H = -2 \sum_{\seq{jk}} J_{jk} \Pi_{jk} \,,
\end{equation}
where we choose a convention that $\Pi_{jk}$ projects onto the trivial state $1$ of Eq.~\eqref{Eq:FusionRule}.
In analogy
with ordinary spin systems we refer to $J_{jk} > 0$ as antiferromagnetic (AFM) interactions and $J_{jk} < 0$ as
ferromagnetic (FM) interactions.

The microscopics governing the interaction strength $J_{jk}$ are not universal and in particular depend on
the underlying physical system. For the $\nu=5/2$ quantum Hall state it has been shown \cite{PhysRevLett.103.076801}
that the interactions between quasiholes (qh) exponentially decay with the separation between the qhs
(on the scale of the magnetic length) and oscillate in sign (analogous to RKKY
interactions between magnetic moments). Similar behavior, though quantitatively different in the sign and strength of the
oscillations, was found
for vortex interactions in p$_x$+ip$_y$ superconductors \cite{PhysRevLett.103.107001} and in the
Kitaev honeycomb model \cite{KitaevModelInteractions}.

In the presence of spatial disorder, the Wigner crystal does not form an ideal triangular lattice but
rather exhibits a random distribution of bond lengths. If the variation of these bond lengths exceeds the
characteristic length of the sign oscillations (the magnetic length of the quantum Hall state), the
exchange coupling constants $J_{jk}$ will exhibit significant sign disorder.
Due to the exponential decay of the coupling strength with distance, the essential physics of the anyon-model
with random interactions is expected to be captured by a Hamiltonian of the form \eqref{eq:ham_lattice}
with independently distributed nearest-neighbor exchange interactions $J_{jk}$ of random sign.
Recalling the analogy of the anyonic degrees of freedom to ordinary magnetic moments, the random
anyon problem thus bears a striking similarity with the Edwards-Anderson quantum spin glass
\cite{EdwardsAnderson}.

One natural angle to study the random anyon model is to pursue a strong-randomness
renormalization group (RG) approach \cite{StrongCouplingRG}. However, as we will report
elsewhere \cite{Unpublished}, the system flows towards {\sl weaker} disorder under
this RG and thus is not governed by an infinite randomness fixed point. This
renders the strong-randomness RG approach incapable of capturing the essential physics of
the problem.
Below, we therefore follow an alternative path exploiting the equivalence of our random  anyon model
with the problem of Anderson localization of disordered Majorana fermions. This problem is accessible to exact diagonalization studies
of large samples which allow us to identify the analytic low energy theory for the underlying phase.


\begin{figure}[t]
	\centering
		\includegraphics[width=.9\columnwidth]{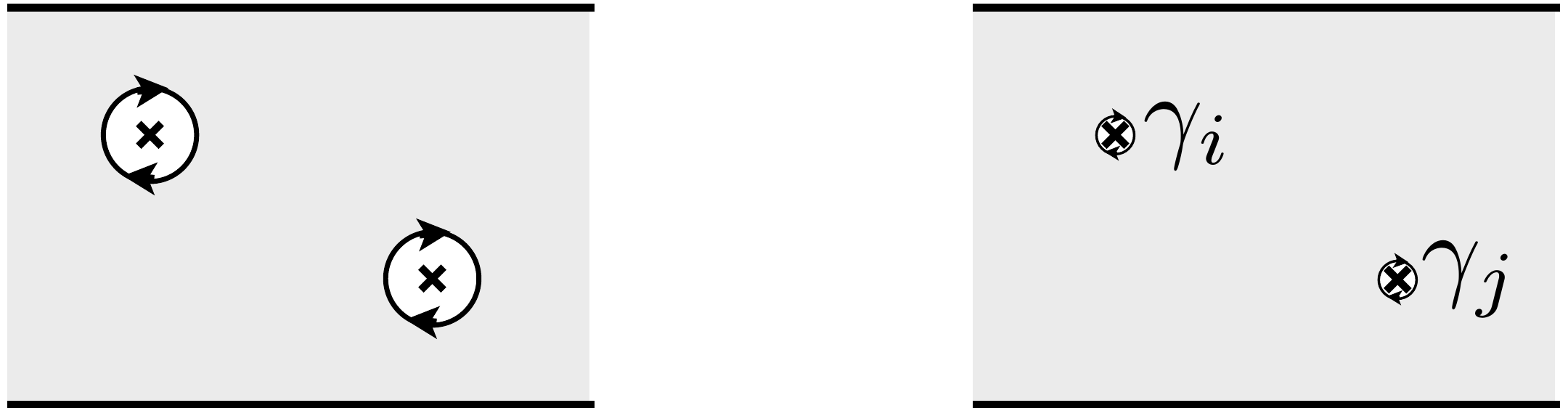}
	\caption{Going from Ising anyons (left) to Majoranas (right).
	             }
	\label{fig:majoranas}
\end{figure}

\paragraph{Majorana fermions.--}
For the $\nu=5/2$ quantum Hall state, quasihole (or quasiparticle) excitations are Ising anyons, which can also be thought of as  vortex-like
excitations carrying edge degrees of freedom as illustrated in Fig.~\ref{fig:majoranas}. The non-Abelian aspects of the Ising anyon $\sigma$ can be described in terms of the Majorana fermion field living on this edge.
The Majorana fermion on an isolated vortex has an exact eigenstate
at zero energy.  We denote the creation operator for such a Majorana
fermion zero-mode by $\gamma_j$. This represents the Ising anyon at position $j$.
In terms of these Majorana operators the original anyon Hamiltonian \eqref{eq:ham_lattice}
can be recast into the form
\begin{equation}
	\label{eq:ham_majorana}
	\mathcal{H} = \sum_{\seq{jk}} i \JJ_{jk} \gamma_j \gamma_{k}
\end{equation}
up to an additive constant.
Because  Majorana fermions are their own Hermitian conjugates, i.e. $\gamma^{\dagger}_j = \gamma^{\phantom\dagger}_j$,
we can interpret Hamiltonian \eqref{eq:ham_majorana} as a hopping problem where the matrix $\JJ_{jk}$ must be real and
antisymmetric in order to ensure Hermiticity \cite{FootnoteGauge}. 
Thus the interacting anyon problem reduces to a hopping problem of non-interacting (Majorana)
fermions \cite{GruzbergLudwigReadNishimori}.
This mapping was used to solve the {\em uniform} interacting anyon problem on a triangular lattice \cite{PhysRevB.73.201303},
which maps to a hopping problem with uniform $\pi/2$ flux around each triangular plaquette, producing a gapped topological state with non-vanishing Chern number.
This was later understood to be a special case of topological liquid nucleation \cite{PhysRevLett.103.070401,LiquidsPaper}
in interacting non-Abelian systems.
In general, two \emph{distinct} gapped topological phases emerge depending on the sign of the (uniform) coupling $J$ in any given non-Abelian theory \cite{PhysRevLett.103.070401,LiquidsPaper}.
For the case of Majorana fermion zero-modes, the nucleated topological liquids are both Abelian
and if brought into spatial proximity would be separated by an edge
state of conformal central charge $c=1$, which corresponds to a difference $\Delta \nu = 2$ in Chern numbers.

\begin{figure}[t]
	\centering
		\includegraphics[width=\columnwidth]{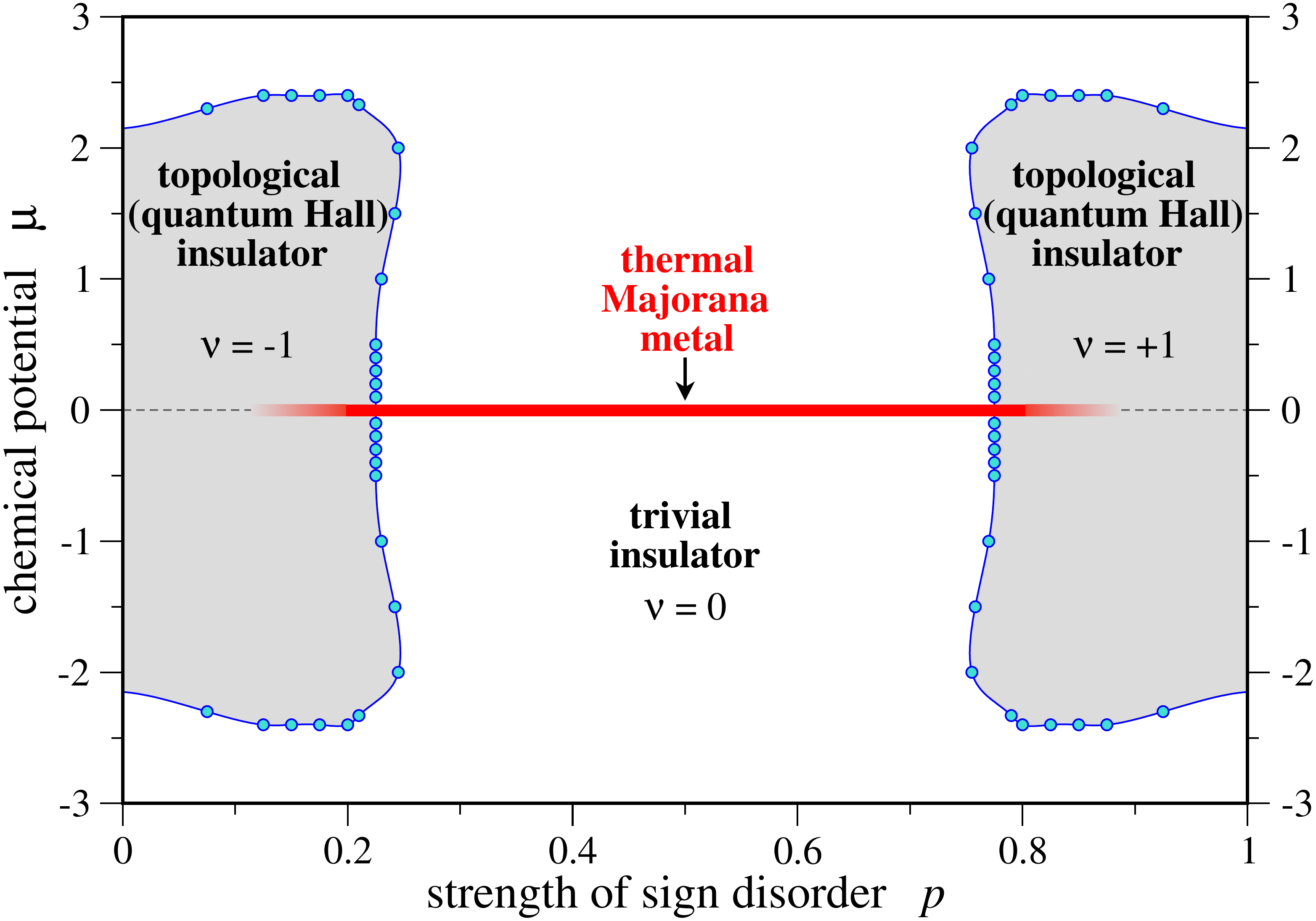}
	\caption{Phase diagram of sign-disordered triangular lattice hopping problem in the $(p,\mu)$ plane. }
	\label{fig:figs_pd}
\end{figure}

{\it In the presence of disorder} we are interested in the physics of anyons  with interactions
of {\it random signs}.
We introduce sign disorder \cite{FootnoteRandomness} by randomly flipping the sign of the couplings $J_{jk}= \pm 1$ on the lattice
independently with probability $p \in [0,1]$.
The limits $p=0$ and $p=1$ then correspond to the uniform problem with $J=+1$ and $J=-1$,
for which the ground states are gapped topological states, which can be characterized by Chern numbers $\nu=-1$
and $\nu=+1$, respectively.
Around these limits these topologically-insulating ground states appear to be stable to a small amount of sign disorder, as indicated in the phase diagram of Fig.~\ref{fig:figs_pd}.  They are gapped only in the limit of strictly zero disorder, while for weak disorder there should be a low density of Anderson-localized states for energies $|E|$ less than a critical energy.
We note that the phase diagram is symmetric under $p \to 1-p$ followed by time reversal of the Majorana fermions,
which is a (statistical) symmetry of the ensemble of Hamiltonians \eqref{eq:ham_majorana}.

\begin{figure}[t]
	\centering
		\includegraphics[width=\columnwidth]{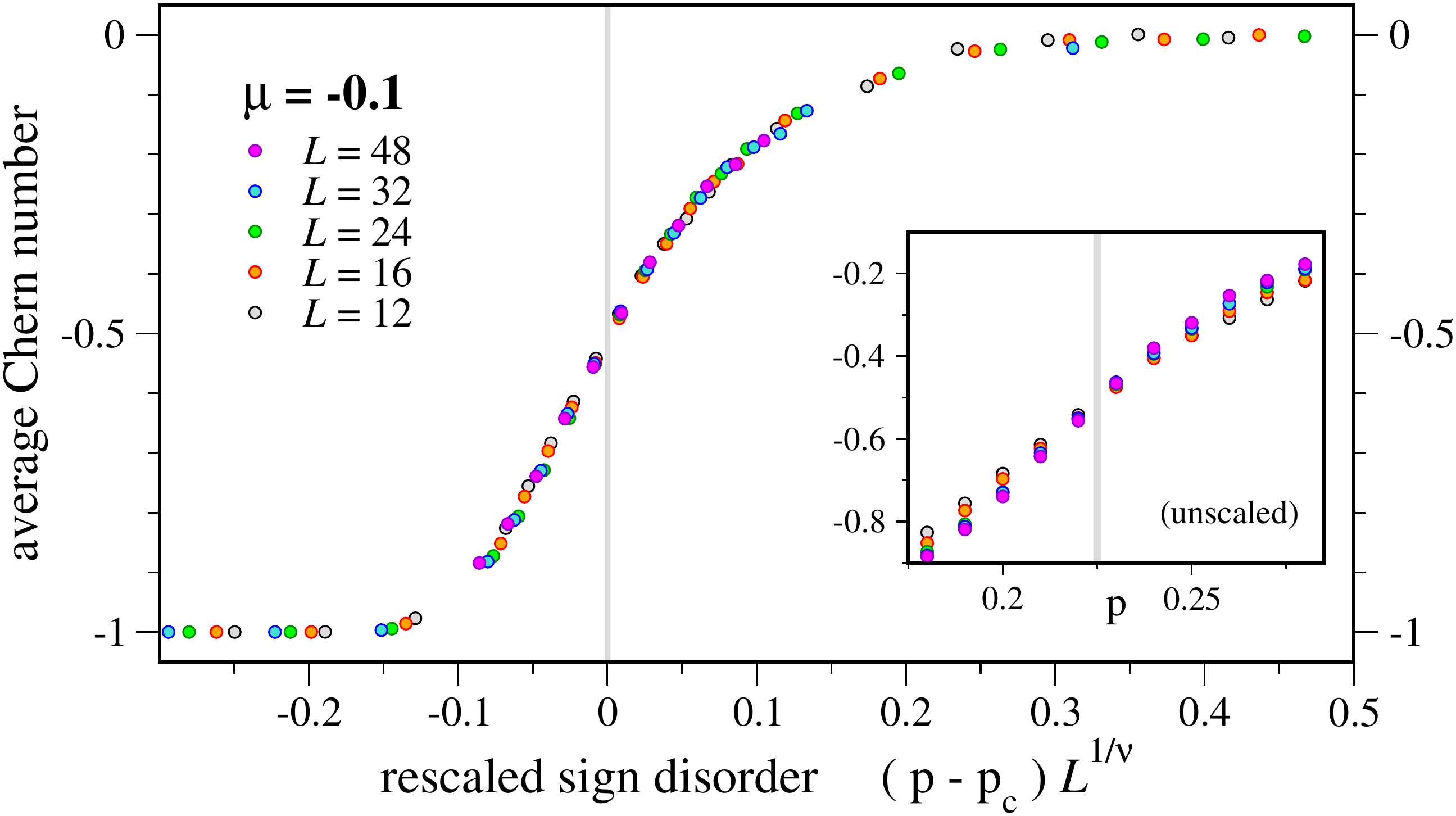}
	\caption{Finite-size scaling collapse of the disorder averaged Chern number $\bar{\nu}$ at $\mu=-0.1$
	              with $p_c \approx 0.225$ and $\nu \approx 5\pm1$.}
	\label{fig:chern-e1}
\end{figure}
\begin{figure}[t]
	\centering
		\includegraphics[width=\columnwidth]{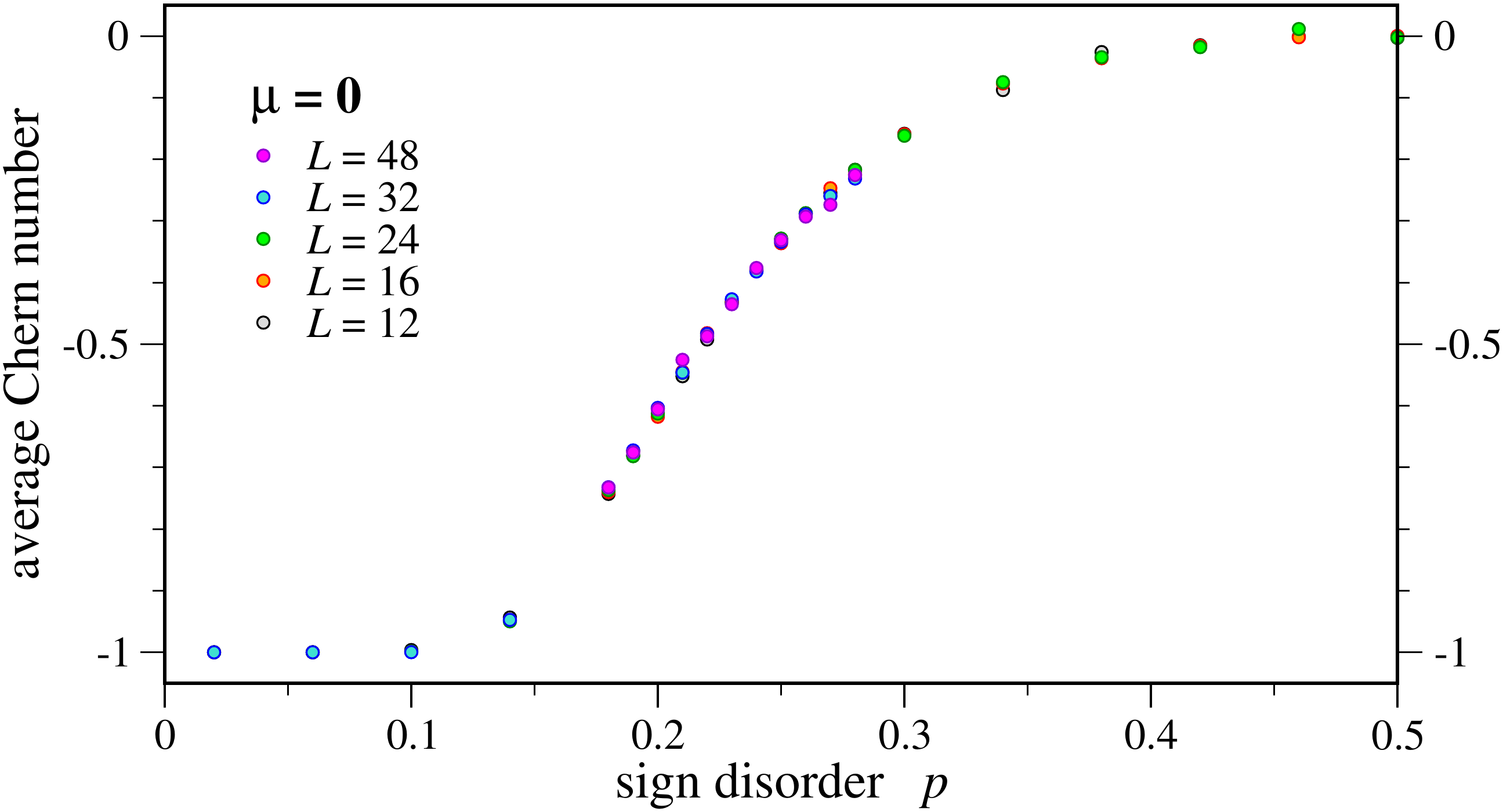}
	\caption{Crossover of the disorder averaged Chern number $\bar{\nu}$ at $\mu=0$ with no visible finite-size scaling.}
	\label{fig:chern-e0}
\end{figure}

In order to determine what happens at intermediate disorder, we numerically investigate the fermion hopping problem \eqref{eq:ham_majorana}.
To this end, we  diagonalize the matrix $i \JJ_{jk}$ from Eq.~\eqref{eq:ham_majorana} by exact diagonalization techniques for lattices ranging
from $L\times L = 8\times 8$ to $64\times 64$ and considering $10^3-10^4$ disorder realizations for various disorder strengths $p$.
For each disorder realization we calculate \cite{Loring:2010p11171} the Chern invariant $\nu$ for eigenstates filled up to a chemical
potential $\mu$.
While it is beneficial to consider the phase diagram in this enlarged parameter space $(p,\mu)$,
we note that the physically relevant ground state of Hamiltonian \eqref{eq:ham_majorana} corresponds to the line
of precisely $\mu = 0$.
Going  to non-zero $\mu$ first, we find a clear transition from a topological insulator with $\nu=-1$ to a topologically trivial phase with $\nu=0$
as $p$ is increased from $0$ to $1/2$.
We can estimate the critical coupling $p_c(\mu)$ for this transition via a finite-size scaling collapse  of the disorder averaged Chern number
$\bar{\nu}$, as shown in Fig.~\ref{fig:chern-e1}. These estimates provide the indicated phase boundary of the topological insulator
in the phase diagram of Fig.~\ref{fig:figs_pd}.
At $\mu=0$, the chemical potential relevant to the original Majorana problem \eqref{eq:ham_majorana}, the disorder averaged Chern number $\bar{\nu}$
shows a slow crossover from $\nu=-1$ in the topological insulator to a state with $\nu=0$ around $p=1/2$ with no visible finite-size scaling, as shown in Fig.~\ref{fig:chern-e0}. 
This state with $\nu=0$ at vanishing chemical potential $\mu=0$ is not the trivial insulator seen at finite chemical potential,
but rather a metallic state as we will demonstrate in the following.

\begin{figure}[b]
	\centering
		\includegraphics[width=\columnwidth]{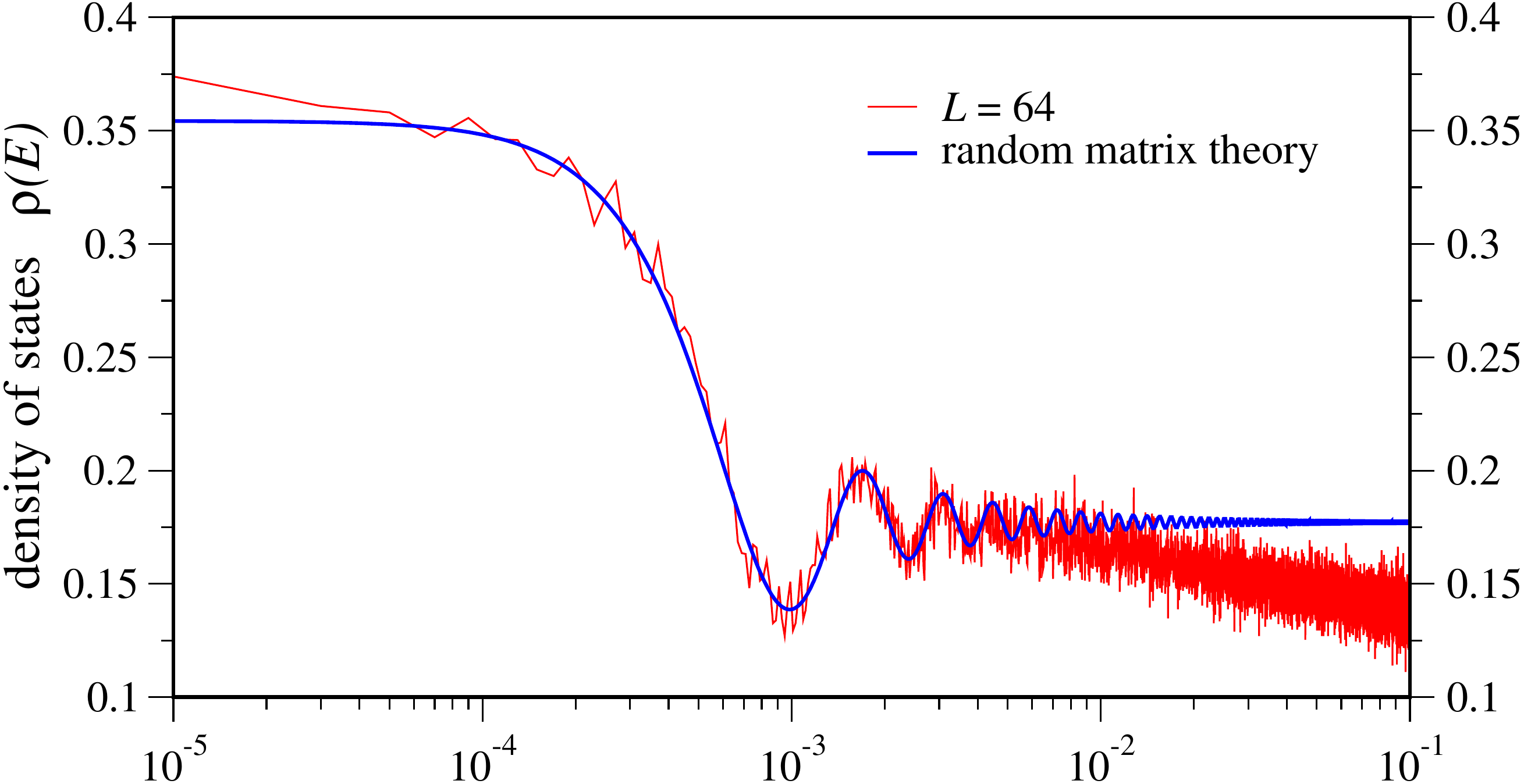}
		\vskip 1mm
		\includegraphics[width=\columnwidth]{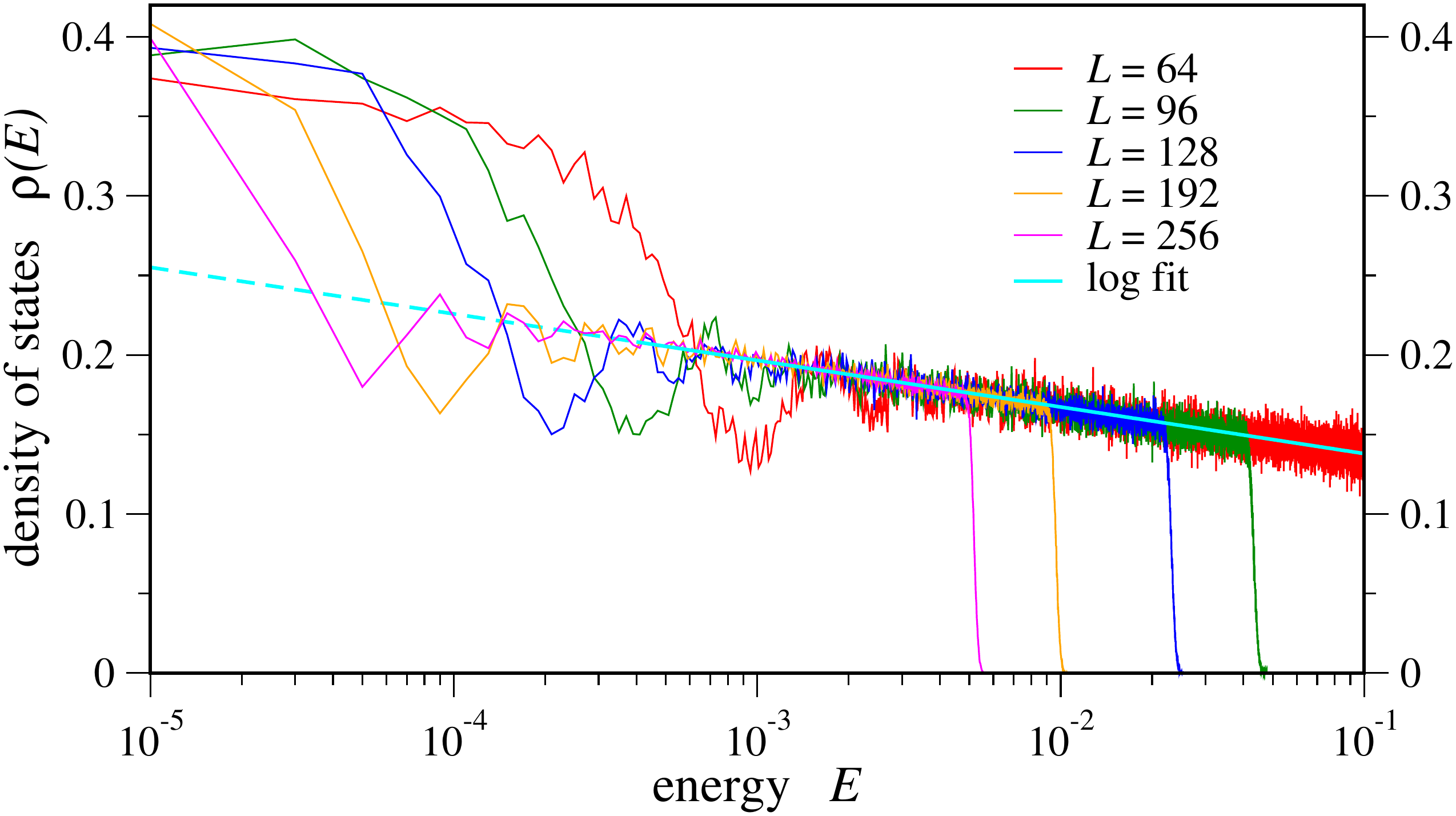}
	\caption{Disorder averaged density of states $\rho(E)$ near $E=0$ at $p=1/2$ collected from $N=2-3\times10^4$ samples per size $L$.}
	\label{fig:dos}
\end{figure}

\paragraph{The thermal metal.--}
In the language of Anderson localization, the random Majorana hopping problem \eqref{eq:ham_majorana} falls into what is known as symmetry class D~\cite{AltlandZirnbauer,Zirnbauer}.
It has been shown~\cite{GruzbergLudwigReadNishimori,ClassDMetallicPhase,ChalkerEtAlClassD}
that problems in this symmetry class can exhibit three distinct phases:
a topological insulator (or thermal quantum Hall insulator), a (topologically) trivial insulator and a {\em thermal metal} phase.
The most remarkable of these three phases, the thermal metal phase,
has previously  been found
only within the framework of Chalker-Coddington network models \cite{Chalker:2001,ChalkerEtAlClassD},
which are not directly related to a microscopic situation.
We will show that the $\nu=0$ state occurring at vanishing $\mu=0$ in our random Majorana hopping problem around $p=1/2$ is precisely
this thermal metal phase \cite{FootnoteNetworkModels}, thereby also providing the first concrete microscopic situation for the appearance of this metallic state.

The most direct evidence for this metallic state comes from the density of eigenstates of Hamiltonian \eqref{eq:ham_majorana}
near zero energy $E=0$ obtained in the numerical solution. We have calculated disorder averages of the density of states $\rho(E)$ using standard sparse diagonalization techniques for system sizes up to $256\times 256$, again with $\sim 2\times10^4$ disorder realizations for a given value of disorder strength $p$.
These are plotted on a semilog scale in Fig.~\ref{fig:dos} for $p=0.5$, clearly revealing
a logarithmic divergence characteristic of the predicted behavior for the thermal metal phase
in class D~\cite{SenthilFisher}. Further, the bump and oscillations visible at the lowest energies (of the order of the inverse mean level spacing) directly follow the form predicted from random matrix theory \cite{AltlandZirnbauer}
$
    \rho(E) = \alpha + {\sin(2\pi \alpha E L^2)}/{(2\pi E L^2)} \,,
$
where $\alpha$ is a non-universal parameter.

\begin{figure}[t]
	\centering
		\includegraphics[width=\columnwidth]{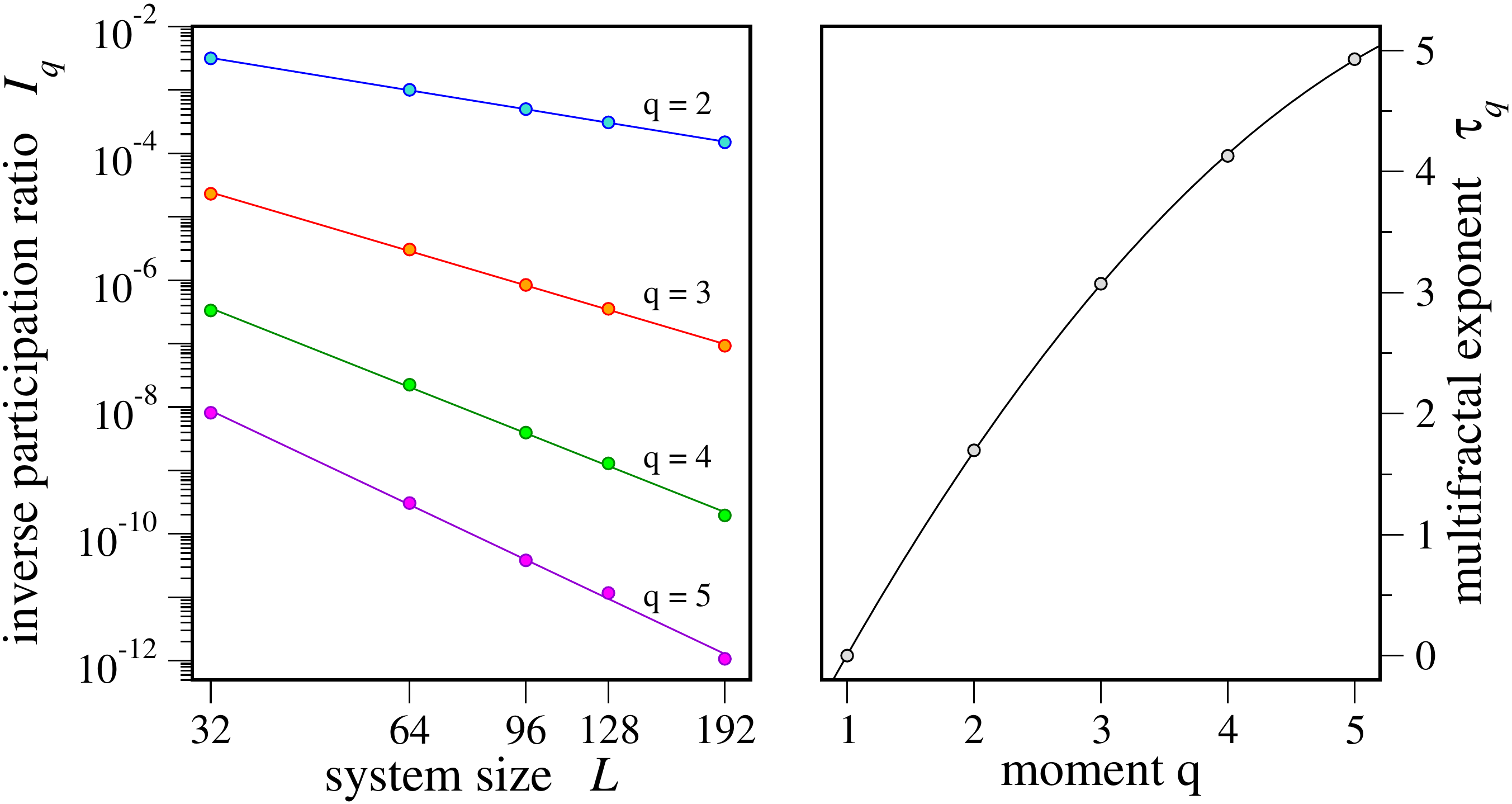}
	\caption{Left: The inverse participation ratios of the wavefunction for disorder strength $p=0.5$
	                        averaged over $N=2\times10^3$ samples per size $L$ in an energy window $E < 10^{-3}$.
	              Right: The multifractal exponents characterizing the decay of the inverse participation ratios.}
	\label{fig:wavefunction}
\end{figure}

Further evidence of the metallic state reveals itself upon close inspection of the wavefunctions near $E=0$.
A metallic state in symmetry class D has a weak antilocalization correction corresponding to a marginally
irrelevant perturbation \cite{SenthilFisher}, in the RG sense.
This implies that the moments of the wavefunctions, or the so-called inverse participation ratios,
exhibit `weak multifractal' behavior (for the $q$-th moment) \cite{MultiFractal}
\[
     I_q = \int d^2r \, | \psi(r) |^{2q} \sim \frac{1}{L^{\tau_q}} \,,
\]
with multifractal exponent $\tau_q = 2(q-1) - \gamma q(q-1)$, where the non-universal coefficient $\gamma$
depends only logarithmically on the linear system size $L$.
The moments of our numerically obtained wavefunctions precisely follow this predicted decay with system size
as shown in the left panel of Fig.~\ref{fig:wavefunction}. The extracted multifractal exponents for different moments
$q$ are fitted by $\tau_q = -0.151q^2 + 2.135q - 1.979$, which  matches the predicted one-parameter form
with $\gamma = 0.151$. 
Finally, we note that the thermal metal 
is robust, in particular with respect to potential higher-order interactions between
the Majorana zero modes $\gamma_i$ in Eq.~\eqref{eq:ham_majorana}.
These have been found to be irrelevant (in the RG sense) in the metallic phase \cite{LudwigSenthilJengChamon}.


\paragraph{Conclusions.--}
To summarize our results, we have demonstrated that moderate disorder induces a thermal metal phase
in systems which harbor a finite density of interacting Majorana fermion zero modes. A prominent candidate
in which to look for this thermal metal phase may be the $\nu=5/2$ quantum Hall plateau, which has long been suggested
to exhibit non-Abelian anyon excitations.
In fact, the experimental observation of thermal heat transport in the bulk may help to reveal the existence of this
non-Abelian state. In abelian states, the thermal conductivity (divided by temperature) would vanish when approaching zero temperature, in contrast to the logarithmic divergence exhibited by the disordered thermal metal \cite{SenthilFisher}. 
The occurrence of this thermal metal state is not restricted to the $\nu=5/2$ quantum Hall state, but is expected
to be present in any two-dimensional time reversal breaking topological phase supporting pinned Majorana zero modes in the presence of moderate disorder -- this includes, e.g., p$_x$+ip$_y$ superconductors
and microscopic realizations of Kitaev's honeycomb model.


\paragraph{Acknowledgments.--}
We thank G. Refael, S. Ryu, and M. Troyer for collaboration on closely related projects and P. Bonderson for stimulating discussions. A.W.W. L. was supported, in part, by NSF DMR-0706140. D.A.H. was supported, in part, by NSF DMR-0819860. C.R.L. acknowledges support from a Lawrence Gollub fellowship and the NSF through a grant for the Institute for Theoretical Atomic and Molecular Physics (ITAMP) at Harvard University.



\begin{thebibliography}{99}
\bibitem{FootnotePairwiseInteractions}
We note here a simplifying feature of the Ising anyons: any multiparty virtual braiding process
involving more than two $\sigma$ anyons may be reduced to a sum of pairwise terms \cite{Parsa}.
\bibitem{FootnoteGauge}
In defining $\JJ_{jk}$ one must make an appropriate choice of $\mathbb{Z}_2$ gauge for the signs of the hoppings. The hopping magnitude $|\JJ_{jk}|$ is given by $|J_{jk}|$.
\bibitem{FootnoteRandomness}
Due to universality one expects the results independent of the chosen short-range distribution for the random variable $J_{jk}$.
This is also explicitly born out by our numerics.
\end{thebibliography}
\end{document}